\documentclass[12pt]{article}
\usepackage{a4}
\usepackage{amsfonts,amssymb,amsmath}
\usepackage{graphicx}
\usepackage{cite}

\begin{document}

\title{Dynamical noncommutativity}
\author{M. Gomes, V.G. Kupriyanov, A.J. da Silva \\
\\
Instituto de F\'{\i}sica, Universidade de S\~{a}o Paulo, Brazil}
\date{\today }
\maketitle

\begin{abstract}
The model of dynamical noncommutativity is proposed. The system consists of
two interrelated parts. The first of them describes the physical degrees of
freedom with coordinates $q^{1}$, $q^{2}$, the second one corresponds to the
noncommutativity $\eta$ which has a proper dynamics. After quantization the
commutator of two physical coordinates is proportional to the function of $\eta$%
. The interesting feature of our model is the dependence of nonlocality on
the energy of the system. The more the energy, the more the nonlocality. The
leading contribution is due to the mode of noncommutativity, however, the
physical degrees of freedom also contribute in nonlocality in higher orders
in $\theta $.
\end{abstract}

\section{Introduction}

It is generally believed, that due to the effects of the gravitational
backreaction the classical background space-time is nonlocal on very small
distances, those comparable with the Planck scale \cite{Doplicher}. Since
noncommutativity usually leads to nonlocality, noncommutative geometry was
proposed as a mathematical tool to systematically incorporate nonlocality in
physics. The simplest approximation to a noncommutative space-time is a flat
canonical noncommutativity, which can be realized by the coordinate
operators $\hat{q}^{\mu }$ satisfying commutation relations $\left[ \hat{q}%
^{\mu },\hat{q}^{\nu }\right] =i\theta ^{\mu \nu }\,,$ where $\theta ^{\mu
\nu \,}$ is an antisymmetric constant matrix. Thus, quantum field theory,
QFT, and quantum mechanics, QM, on flat noncommutative spaces have recently
been extensively studied, see e.g. \cite{NCreviews} and \cite{NCQM} for
reviews on noncommutativity in QFT and QM, respectively.

However, it is not reasonable to suppose that the noncommutativity and hence
nonlocality is the same in the whole universe. It can depend on the point of
the space-time, see e.g., \cite{Szabo} and references therein. A model of position-dependent
noncommutativity in quantum mechanics, involving the complete algebra of noncommutative coordinates:%
\begin{equation}
\left[ \hat{q}^{\mu },\hat{q}^{\nu }\right] =i\omega ^{\mu \nu }\left( \hat{q%
}\right) ~,  \label{3}
\end{equation}%
where $\omega ^{\mu \nu }$ is some given matrix-valued function, and corresponding momenta was recently proposed in \cite{GK09}.

To precisely specify $\omega ^{\mu \nu }\left( q\right) $, physical
considerations must be taken into account. Strong magnetic fields, quantum
oscillations associated to pair creation and strings dynamics are just some
causes that may lead to noncommutativity. A basic question here is to
uncover the mechanism leading to noncommutativity. One of the possibilities
is to suppose that noncommutativity has its own degrees of freedom which
interact in some way with the physical degrees of freedom of the system.
Since noncommutativity $\omega ^{\mu \nu }\left( q\right) $ is a field it
has an infinite number of degrees of freedom and its evolution should be
described by a field equation of motion. To simplify our considerations and
to better understand what happens, we suppose that noncommutativity has only
one degree of freedom, embodied by the variable $\eta$. We also suppose that
the physical system under study is described by two coordinates $q^{1}$ and $%
q^{2}$. In this situation, the most general form of the commutation
relations is
\begin{equation}
\left[ \hat{q}^{i},\hat{q}^{j}\right] =i\theta f\left( \hat{\eta}\right)
\varepsilon ^{ij},  \label{4}
\end{equation}%
where $f$ is some function of the indicated argument and $\theta $ is a
parameter which gives the strength of the noncommutativity. By itself,
Equation (\ref{4}) provides the main effect of the interaction between $\eta$
and $q_{1}$, $q_{2}$. This type of commutation
relation we call dynamical noncommutativity.

In this work, we consider a classical model which after quantization
reproduces the dynamical noncommutativity (\ref{4}) and discuss some of its
properties. An interesting feature of the system in question is the
dependence of the nonlocality on the energy of the system. The leading
contribution comes from the modes associtated to the variable $\eta$, but the
physical degrees of freedom $q_{1}$ and $q_{2}$ also contribute to
nonlocality in higher orders.

\section{The classical model and quantization}

Consider the following singular Lagrangian\footnote{%
For simplicity we consider two-dimensional case, $q^{i}=(q^{1},q^{2})$,
however, the generalization to the $n$-dimensional case is straightforward.}:%
\begin{eqnarray}
L &=&L_{q}+L_{r}+L_{int}~,  \label{1} \\
L_{q} &=&v_{i}\dot{q}^{i}-h\left( v,q\right) ,\ \ L_{\eta}=\frac{\dot{\eta}^{2}}{2}%
-V\left( \eta\right) ,\ L_{int}=\frac{\theta }{2}f\left( \eta\right) \varepsilon
^{ij}v_{i}\dot{v}_{j},\   \notag
\end{eqnarray}%
here $h\left( v,q\right) $ and $V\left( \eta\right) $ are some given functions
and $q,,\ v,,\ \eta$ are an independent variables\footnote{One can also consider the situation when $v^{i}$ are velocities, $v^{i}=\dot{q}^{i}$, i.e., a model with higher derivatives, similar to the one proposed in \cite{Lukierski}. This model has some difficulties with the physical interpretation of the additional degrees of freedom, also, its quantum spectrum is not bounded from below \cite{HP}.}
It consists of three parts: $L_{q}$ and $L_{\eta}$ describe dynamics of
two subsystems with coordinates $q^{1},\ q^{2}$ and $\eta$ correspondingly, and
$L_{int}$ describes the interaction between these two subsystems. Notice
that first order actions containing a term similar to $L_{int}$ appear in the context of the
Landau problem in the limit of strong magnetic field or small electron mass \cite{Jackiw} and lead to
noncommutativity between the coordinates $q^{i}$ after quantization, see e.g.
\cite{Kup}. Actually, $L_{int}$ is the minimal interaction needed to produce
the desired commutation relation (\ref{4}).

To proceed to the Hamiltonian formulation, we introduce conjugate momenta
according to the rules%
\begin{equation}
p_{i}=\frac{\partial L}{\partial \dot{q}^{i}},\ \ \pi _{i}=\frac{\partial L}{%
\partial \dot{v}^{i}},\ \ p_{\eta}=\frac{\partial L}{\partial \dot{\eta}}~.
\label{2}
\end{equation}%
They are considered as algebraic equations for determining the velocities $%
\dot{q}^{i},\dot{v}^{i},\dot{\eta}.$ Since the Lagrangian (\ref{1}) is linear
in $\dot{q}^{i}$ and $\dot{v}^{i},$ the first two equations of (\ref{2})
become primary constraints in the Hamiltonian formalism:%
\begin{equation}
\ G_{i}=p_{i}-v_{i}=0~,\ \ T_{i}=\pi _{i}+\frac{\theta }{2}f\left( \eta\right)
\varepsilon ^{ij}v_{j}=0~.  \label{8}
\end{equation}%
The total Hamiltonian of the theory is:%
\begin{equation}
H^{\left( 1\right) }=H+\lambda _{p}^{i}G_{i}+\lambda _{\pi }^{i}T_{i},\ \
H=h\left( p,q\right) +\frac{p_{\eta}^{2}}{2}+V\left( \eta\right) ,
\end{equation}%
where $\lambda _{p}^{i}$ and $\lambda _{\pi }^{i}$ are Lagrangian
multipliers. In Dirac's terminology (\ref{8}) are second-class constraints.
Thus, from the consistency conditions for the primary constraints
\begin{equation}
\dot{\Phi}=\{\Phi ,H^{\left( 1\right) }\} =0\,,
\end{equation}%
where $\Phi =\left( G,T\right)$, one can fix all the $\lambda $-s.
Secondary constraints do not appear. The
Hamiltonian equations are%
\begin{equation}
\dot{\chi}=\{\chi ,H\}_{D(\Phi )}\,,\;\Phi =0\,,  \label{10}
\end{equation}%
where $\chi =(q,v,\eta,p,\pi ,p_{\eta})$ and $\{\cdot
,\cdot \}_{D(\Phi )}$ are Dirac brackets with respect to second-class
constraints (\ref{8}).

Since the effective Hamiltonian $H$ depends only on the coordinates $q^{i},\eta$
and momenta $p_{i},p_{\eta}$, we write the Dirac brackets only between these
variables:%
\begin{align}
& \left\{ q^{i},q^{j}\right\} _{D\left( \Phi \right) }=f\left( \eta\right)
\theta \varepsilon ^{ij}~,\ \ \left\{ q^{i},p_{j}\right\} _{D\left( \Phi
\right) }=\delta _{j}^{i}~,  \label{11} \\
& \left\{ q^{i},p_{\eta}\right\} _{D\left( \Phi \right) }=-\frac{\theta }{2}%
f^{\prime }\left( \eta\right) \varepsilon ^{ij}p_{j}~,\ \ \left\{
\eta,p_{\eta}\right\} _{D\left( \Phi \right) }=1  \notag \\
& \left\{ \eta,q^{i}\right\} _{D\left( \Phi \right) }=\left\{ \eta,p_{i}\right\}
_{D\left( \Phi \right) }=\left\{ p_{\eta},p_{i}\right\} _{D\left( \Phi \right)
}=\left\{ p_{i},p_{j}\right\} _{D\left( \Phi \right) }=0.  \notag
\end{align}

After quantization, these Dirac brackets will determine the commutation
relations between operators of coordinates $\hat{q}^{i},\ \hat{\eta}$ and the
corresponding operators of momenta $\hat{p}_{i},\ \hat{p}_{\eta}$:%
\begin{align}
& \left[ \hat{q}^{i},\hat{q}^{j}\right] =i\theta f\left( \hat{\eta}\right)
\varepsilon ^{ij}~,\ \ \left[ \hat{q}^{i},\hat{p}_{j}\right] =i\delta
_{j}^{i}~,  \label{12} \\
& \left[ \hat{q}^{i},\hat{p}_{\eta}\right] =-\frac{i\theta }{2}f^{\prime
}\left( \hat{\eta}\right) \varepsilon ^{ij}\hat{p}_{j}~,\ \ \left[ \hat{\eta},\hat{%
p}_{\eta}\right] =i  \notag \\
& \left[ \hat{\eta},\hat{q}^{i}\right] =\left[ \hat{\eta},\hat{p}_{i}\right] =%
\left[ \hat{p}_{\eta},\hat{p}_{i}\right] =\left[ \hat{p}_{i},\hat{p}_{j}\right]
=0.  \notag
\end{align}%
That is, dynamical noncommutativity (\ref{4}) can be reproduced by the
quantization of the system with Lagrangian (\ref{1}). The quantum
Hamiltonian is%
\begin{equation}
\hat{H}=h\left( \hat{p},\hat{q}\right) +\frac{\hat{p}_{\eta}^{2}}{2}+V\left(
\hat{\eta}\right) ~,  \label{13}
\end{equation}%
where some ordering is implicitly assumed.

To construct representation of the operatorial algebra\footnote{%
Note that in the general case, the polydifferential representation of the
operatorial algebra defined by the commutation relations $\left[ \hat{\xi}%
^{i},\hat{\xi}^{j}\right] =i\omega ^{ij}\left( \hat{\xi}\right) ,$ was
constructed in \cite{KV}.} (\ref{12}) let us introduce new operators%
\begin{equation}
\hat{x}^{i}=\hat{q}^{i}+\frac{\theta }{2}f\left( \hat{\eta}\right) \varepsilon
^{ij}\hat{p}_{j}~.  \label{14}
\end{equation}%
It is easy to see that operators $\hat{x}^{i},\ \hat{\eta},\ \hat{p}_{i},\ \hat{%
p}_{\eta}$ obey canonical commutation relations:%
\begin{align}
& \left[ \hat{x}^{i},\hat{p}_{j}\right] =i\delta _{j}^{i}~,\left[ \hat{\eta},%
\hat{p}_{\eta}\right] =i~,  \label{15} \\
& \left[ \hat{x}^{i},\hat{x}^{i}\right] =\left[ \hat{\eta},\hat{x}^{i}\right] =%
\left[ \hat{x}^{i},\hat{p}_{\eta}\right] =0~,~\ \   \notag \\
& \left[ \hat{\eta},\hat{p}_{i}\right] =\left[ \hat{p}_{i},\hat{p}_{j}\right] =%
\left[ \hat{p}_{\eta},\hat{p}_{i}\right] =0~.  \notag
\end{align}%
Thus, similarly to a Bopp shift \cite{Bopp}, we pass to the new set of
operators $\hat{x}^{i},\ \hat{\eta},\ \hat{p}_{i},\ \hat{p}_{\eta}$ with canonical
commutation relations (\ref{15}). The quantum Hamiltonian (\ref{13}) in the
new variables will take the form%
\begin{equation}
\hat{H}=h\left( \hat{p}_{i},\hat{x}^{i}-\frac{\theta }{2}f\left( \hat{\eta}%
\right) \varepsilon ^{ij}\hat{p}_{j}\right) +\frac{\hat{p}_{\eta}^{2}}{2}%
+V\left( \hat{\eta}\right) ~.  \label{16}
\end{equation}%
To investigate possible physical implications of dynamical noncommutativity
it is reasonable to start with exactly solvable models like the harmonic
oscillator and the hydrogen atom and then compare our results with those of
canonical noncommutative quantum mechanics {\cite{Chaichian,Horvathy}}. For
the harmonic oscillator, we suppose that
\begin{equation}
h\left( \hat{p},\hat{q}\right) =\frac{\hat{p}_{i}^{2}}{2}+\frac{\omega
_{i}^{2}\left( \hat{q}^{i}\right) ^{2}}{2},\ \ V\left( \hat{\eta}\right) =\frac{%
\omega _{\eta}^{2}\hat{\eta}^{2}}{2},  \label{17}
\end{equation}%
so that, after the Bopp shift, the Hamiltonian (\ref{13}) takes the form%
\begin{equation}
\hat{H}=\frac{\hat{p}_{i}^{2}}{2}+\frac{\omega _{i}^{2}}{2}\left( \hat{x}%
^{i}-\frac{\theta }{2}f\left( \hat{\eta}\right) \varepsilon ^{ij}\hat{p}%
_{j}\right) ^{2}+\frac{\hat{p}_{\eta}^{2}}{2}+\frac{\omega _{\eta}^{2}}{2}\hat{\eta}%
^{2}~.  \label{18}
\end{equation}

In the coordinate representation, the operators $\hat{x}^{i}=x^{i}\cdot $
and $\hat{\eta}=\eta\cdot $ are multiplicative whereas the momenta operators act
as derivatives: $\hat{p}_{i}=-i\partial _{i}$ and $\hat{p}_{\eta}=-i\partial _{\eta}$
We employ a perturbation theory over the parameter of noncommutativity $%
\theta $, starting with the quantum mechanical problem described by the
Hamiltonian (\ref{16}) with $\theta =0$ and then construct corrections for $%
\theta \neq 0$.

\section{Nonlocality caused by dynamical noncommutativity}

It would be interesting to calculate the uncertainty relations between the
operators of the noncommutative coordinates $\hat{q}^{i}$, i.e., the
nonlocality caused by the dynamical noncommutativity. According to (\ref{12}%
) one has:%
\begin{equation}
\left( \bigtriangleup q^{1}\right) _{\Psi }\left( \bigtriangleup
q^{2}\right) _{\Psi }\geq \frac{\theta }{2}\left\vert \left\langle \Psi
\right\vert f\left( \hat{\eta}\right) \left\vert \Psi \right\rangle \right\vert
,  \label{20}
\end{equation}%
where $\left\vert \Psi \right\rangle $ is a given state. If
we choose $\left\vert \Psi \right\rangle $ as a stationary state of the system, i.e., an eigenstate
of the Hamiltonian (\ref{18}), it is reasonable to expect that nonlocality
will depend on the energy. In general, this is true, however,
in some specific models, e.g., $f(\eta)=\eta$ such dependence does not appear.

To calculate the uncertainty relations (\ref{20}) and the energy spectrum
for the quantum system with the Hamiltonian (\ref{18}) we employ the
standard perturbation procedure (see e.g., \cite{LL}). Note, however, that
the result of this calculation will not depend neither on the choice of basis
of operators, nor on the perturbation method used to solve the problem. First, the
hamiltonian (\ref{18}) is separated into two parts as%
\begin{equation}
\hat{H}=\hat{H}_{0}+\hat{V},  \label{21}
\end{equation}%
where $\hat{H}_{0}$ is the Hamiltonian of the harmonic oscillator:%
\begin{equation}
\hat{H}_{0}=\frac{\hat{p}_{i}^{2}}{2}+\frac{\omega _{i}^{2}}{2}\left( \hat{x}%
^{i}\right) ^{2}+\frac{\hat{p}_{\eta}^{2}}{2}+\frac{\omega _{\eta}^{2}}{2}\hat{\eta}%
^{2}~,  \label{22}
\end{equation}%
with eigenstates (we use here the notations of creation and annihilation
operators, $\left\vert n\right\rangle =\frac{\left( a^{+}\right) ^{n}}{\sqrt{%
n!}}\left\vert 0\right\rangle $):%
\begin{equation}
\left\vert \Psi _{lmn}^{\left( 0\right) }\right\rangle =\left\vert
l\right\rangle \otimes \left\vert m\right\rangle \otimes \left\vert
n\right\rangle ,  \label{23}
\end{equation}%
and eigenvalues%
\begin{equation}
E_{lmn}^{\left( 0\right) }=\omega _{1}\left( l+\frac{1}{2}\right) +\omega
_{2}\left( m+\frac{1}{2}\right) +\omega _{\eta}\left( n+\frac{1}{2}\right) ,
\label{24}
\end{equation}%
and%
\begin{eqnarray}
&&\hat{V}=\hat{V}_{1}+\hat{V}_{2},  \label{25} \\
&&\hat{V}_{1}=\frac{\theta }{2}f\left( \hat{\eta}\right) \left( \omega _{1}^{2}%
\hat{x}^{2}\hat{p}_{1}-\omega _{2}^{2}\hat{x}^{1}\hat{p}_{2}\right) ,\ \
\hat{V}_{2}=\frac{\theta ^{2}\omega _{i}^{2}\hat{p}_{i}^{2}}{8}f\left( \hat{\eta%
}\right) ^{2},  \notag
\end{eqnarray}%
is the perturbation. We assume that the set $\left\vert \Psi _{lmn}^{\left(
0\right) }\right\rangle $ provides an orthonormal basis in the Hilbert space
of the system, i.e.,%
\begin{equation}
\left\langle \Psi _{lmn}^{\left( 0\right) }\right\vert \left. \Psi
_{l^{\prime }m^{\prime }n^{\prime }}^{\left( 0\right) }\right\rangle =\delta
_{ll^{\prime }}\delta _{mm^{\prime }}\delta _{nn^{\prime }},\ \ I=\underset{%
lmn}{\sum }\left\vert \Psi _{lmn}^{\left( 0\right) }\right\rangle
\left\langle \Psi _{lmn}^{\left( 0\right) }\right\vert .  \label{26}
\end{equation}%
To solve the eigenvalue problem for (\ref{21}), we write:%
\begin{eqnarray}
&&\hat{H}\left\vert \Psi \right\rangle =E\left\vert \Psi \right\rangle ,
\label{28} \\
&&\left\vert \Psi \right\rangle =\left\vert \Psi _{lmn}\right\rangle
=\left\vert \Psi _{lmn}^{\left( 0\right) }\right\rangle +\left\vert \Psi
_{lmn}^{\left( 1\right) }\right\rangle +...,\ \ E=E^{\left( 0\right)
}+E^{\left( 1\right) }+...,  \notag
\end{eqnarray}%
and express the eigenfunction $\Psi $ in terms of the eigenstates of
harmonic oscillator:
\begin{eqnarray}
&&\left\vert \Psi _{lmn}\right\rangle =\underset{l^{\prime }m^{\prime
}n^{\prime }}{\sum }C_{lmn}^{l^{\prime }m^{\prime }n^{\prime }}\left\vert
\Psi _{l^{\prime }m^{\prime }n^{\prime }}^{\left( 0\right) }\right\rangle ~,
\label{29} \\
&&C_{lmn}^{l^{\prime }m^{\prime }n^{\prime }}=C_{lmn}^{\left( 0\right)
l^{\prime }m^{\prime }n^{\prime }}+C_{lmn}^{\left( 1\right) l^{\prime
}m^{\prime }n^{\prime }}+...~,\ \ C_{lmn}^{\left( 0\right) l^{\prime
}m^{\prime }n^{\prime }}=\delta _{l}^{l^{\prime }}\delta _{m}^{m^{\prime
}}\delta _{n}^{n^{\prime }}.  \notag
\end{eqnarray}%
From (\ref{28}) one finds the equation on the coefficients $%
C_{lmn}^{l^{\prime }m^{\prime }n^{\prime }}$ of the decomposition (\ref{29})
and eigenvalues $E_{lmn}$:(here the summation is only over $l^{\prime \prime
}m^{\prime \prime }n^{\prime \prime }$)
\begin{equation}
\left( E_{lmn}-E_{l^{\prime }m^{\prime }n^{\prime}}^{\left( 0\right)
}\right) C_{lmn}^{l^{\prime }m^{\prime }n^{\prime }}=\underset{l^{\prime
\prime }m^{\prime \prime }n^{\prime \prime }}{\sum }V_{l^{\prime \prime
}m^{\prime \prime }n^{\prime \prime }}^{l^{\prime }m^{\prime }n^{\prime
}}C_{lmn}^{l^{\prime \prime }m^{\prime \prime }n^{\prime \prime }}~,
\label{30}
\end{equation}

\begin{equation}
V_{l^{\prime }m^{\prime }n^{\prime }}^{lmn}\equiv \left\langle \Psi
_{lmn}^{\left( 0\right) }\right\vert \hat{V}\left\vert \Psi _{l^{\prime
}m^{\prime }n^{\prime }}^{\left( 0\right) }\right\rangle= V_{l^{\prime
}m^{\prime }n^{\prime }}^{\left( 1\right) lmn}+V_{l^{\prime }m^{\prime
}n^{\prime }}^{\left( 2\right) lmn},  \label{31}
\end{equation}%
where%
\begin{eqnarray}
&&V_{l^{\prime }m^{\prime }n^{\prime }}^{\left( 1\right) lmn} = \label{33} \\ &&\frac{%
i\theta }{4\sqrt{\omega _{1}\omega _{2}}}\left\langle n\right\vert f\left(
\hat{\eta}\right) \left\vert n^{\prime }\right\rangle \left[ \left( \omega
_{1}^{3}-\omega _{2}^{3}\right) \left( \sqrt{l^{\prime }+1}\sqrt{m^{\prime
}+1}\delta _{l,l^{\prime }+1}\delta _{m,m^{\prime }+1}\right. \right.
\notag \\
&&-\left. \sqrt{l^{\prime }m^{\prime }}\delta _{l,l^{\prime }-1}\delta
_{m,m^{\prime }-1}\right) +\left( \omega _{1}^{3}+\omega _{2}^{3}\right)
\left( \sqrt{l^{\prime }+1}\sqrt{m^{\prime }}\delta _{l,l^{\prime }+1}\delta
_{m,m^{\prime }-1}\right.  \notag \\
&&-\left. \left. \sqrt{l^{\prime }}\sqrt{m^{\prime }+1}\delta _{l,l^{\prime
}-1}\delta _{m,m^{\prime }+1}\right) \right] ,  \notag
\end{eqnarray}%
\begin{eqnarray}
V_{l^{\prime }m^{\prime }n^{\prime }}^{\left( 2\right) lmn} &=&-\frac{\theta
^{2}}{16}\left\langle n\right\vert f\left( \hat{\eta}\right) ^{2}\left\vert n^{\prime
}\right\rangle \left[ \omega _{1}^{3}\left( \sqrt{\left( l^{\prime
}+1\right) \left( l^{\prime }+2\right) }\delta _{l,l^{\prime }+2}-\left(
2l^{\prime }+1\right) \delta _{l,l^{\prime }}\right. \right.  \label{34} \\
&&+\left. \sqrt{l^{\prime }\left( l^{\prime }-1\right) }\delta _{l,l^{\prime
}-2}\right) +\omega _{2}^{3}\left( \sqrt{\left( m^{\prime }+1\right) \left(
m^{\prime }+2\right) }\delta _{m,m^{\prime }+2}\right.  \notag \\
&&-\left. \left. \left( 2m^{\prime }+1\right) \delta _{m,m^{\prime }}+\sqrt{%
m^{\prime }\left( m^{\prime }-1\right) }\delta _{m,m^{\prime }-2}\right) %
\right] .  \notag
\end{eqnarray}

Now let us analyze equation (\ref{30}). It can be solved order by order in $%
\theta $. In the $n$-th order in $\theta $ this equation with $l=l^{\prime
},\ m=m^{\prime }$ and $n=n^{\prime }$ gives the $n$-th order correction to
the eigenvalue $E$. For arbitrary $l,l^{\prime },m,m^{\prime },n,n^{\prime }$%
, under condition that $E_{lmn}^{\left( 0\right) }-E_{l^{\prime }m^{\prime
}n^{\prime }}^{\left( 0\right) }\neq 0$, eq. (\ref{30}) determines the $n$%
-th order of the coefficients $C_{lmn}^{l^{\prime }m^{\prime }n^{\prime }}$.
The $n$-th order of the coefficient $C_{lmn}^{lmn}$ can be determined from
the condition of normalization of the state $\left\vert \Psi
_{lmn}\right\rangle $ up to the $n$-th order in $\theta $. Thus, e.g., in
the first order in $\theta $ one has:%
\begin{equation}
E_{lmn}^{\left( 1\right) }=V_{lmn}^{\left( 1\right) lmn}=0,  \label{37}
\end{equation}%
i.e. the first order correction to energy is equal to zero, like in the case
of anharmonic oscillator. The coefficients $C_{lmn}^{l^{\prime }m^{\prime
}n^{\prime }}$ in the first order are:
\begin{eqnarray}
C_{lmn}^{\left( 1\right) l^{\prime }m^{\prime }n^{\prime }} &=&\frac{1}{%
E_{lmn}^{\left( 0\right) }-E_{l^{\prime }m^{\prime }n^{\prime }}^{\left(
0\right) }}V_{lmn}^{\left( 1\right) l^{\prime }m^{\prime }n^{\prime }},\
\forall l^{\prime },m^{\prime },n^{\prime }\neq n,  \label{38} \\
C_{lmn}^{\left( 1\right) l^{\prime }m^{\prime }n} &=&\frac{1}{%
E_{lmn}^{\left( 0\right) }-E_{l^{\prime }m^{\prime }n}^{\left( 0\right) }}%
V_{lmn}^{\left( 1\right) l^{\prime }m^{\prime }n},\ \forall l^{\prime
},m^{\prime }\neq m,  \notag \\
C_{lmn}^{\left( 1\right) l^{\prime }mn} &=&\frac{1}{E_{lmn}^{\left( 0\right)
}-E_{l^{\prime }mn}^{\left( 0\right) }}V_{lmn}^{\left( 1\right) l^{\prime
}mn},\ \forall l^{\prime }\neq l,  \notag
\end{eqnarray}%
and%
\begin{equation}
C_{lmn}^{\left( 1\right) lmn}=0.
\end{equation}%
The second-order corrections to the energy spectrum $E$ and coefficients of
the decomposition (\ref{29}) can be constructed in the same way, but we do
not present here the exact formulas.

Let us now calculate expectation value%
\begin{eqnarray}
\left\langle \Psi _{lmn}\right\vert f\left( \eta\right) \left\vert \Psi
_{lmn}\right\rangle &=&\underset{l^{\prime }m^{\prime }n^{\prime }n^{\prime
\prime }}{\sum }C_{lmn}^{\ast l^{\prime }m^{\prime }n^{\prime
}}C_{lmn}^{l^{\prime }m^{\prime }n^{\prime \prime }}\left\langle n^{\prime
}\right\vert f\left( \eta\right) \left\vert n^{\prime \prime }\right\rangle
=\left\langle n\right\vert f\left( \eta\right) \left\vert n\right\rangle
\label{43} \\
&&+\underset{n^{\prime }}{\sum }\left( C_{lmn}^{\ast \left( 1\right)
lmn^{\prime }}\left\langle n^{\prime }\right\vert f\left( \eta\right)
\left\vert n\right\rangle +\left\langle n\right\vert f\left( r\right)
\left\vert n^{\prime }\right\rangle C_{lmn}^{\left( 1\right) lmn^{\prime
}}\right)  \notag \\
&&+\underset{n^{\prime }}{\sum }\left( C_{lmn}^{\ast \left( 2\right)
lmn^{\prime }}\left\langle n^{\prime }\right\vert f\left( \eta\right)
\left\vert n\right\rangle +\left\langle n\right\vert f\left( r\right)
\left\vert n^{\prime }\right\rangle C_{lmn}^{\left( 2\right) lmn^{\prime
}}\right)  \notag \\
&&+\underset{l^{\prime }m^{\prime }n^{\prime }n^{\prime \prime }}{\sum }%
C_{lmn}^{\ast \left( 1\right) l^{\prime }m^{\prime }n^{\prime
}}C_{lmn}^{\left( 1\right) l^{\prime }m^{\prime }n^{\prime \prime
}}\left\langle n^{\prime }\right\vert f\left( \eta\right) \left\vert n^{\prime
\prime }\right\rangle +O\left( \theta ^{3}\right) .  \notag
\end{eqnarray}%
One can easily verify that the first order of this decomposition is equal to
zero, because of the special form of the coefficients $C_{lmn}^{\left(
1\right) l^{\prime }m^{\prime }n^{\prime }}$. The first nontrivial
contribution may appear in the second order.

To simplify our considerations we suppose that the system is in the ground
state of noncommutativity, $n=0,\ \left\vert \Psi _{E}\right\rangle
=\left\vert \Psi _{lm0}\right\rangle $, and also, $\omega _{\eta}\gg \omega
_{1},\omega _{2}$, i.e., the probability of the system to pass to the first
excited state of noncommutativity ($n>0$) is very small. In what follows we
assume that $n=0$ and take into account only the leading powers of the ratio
$\omega _{i}/\omega _{\eta}$.

For $f\left( \eta\right) =\eta$, the second order contribution in (\ref{43}) is:
\begin{eqnarray*}
&&\sqrt{\frac{n+1}{2\omega _{\eta}}}\left( C_{lmn}^{\ast \left( 2\right)
lmn+1}+C_{lmn}^{\left( 2\right) lmn+1}\right) +\sqrt{\frac{n}{2\omega _{\eta}}}%
\left( C_{lmn}^{\ast \left( 2\right) lmn-1}+C_{lmn}^{\left( 2\right)
lmn-1}\right) \\
&&+\frac{1}{\sqrt{2\omega _{\eta}}}\underset{l^{\prime }m^{\prime }n^{\prime }}{%
\sum }\left( \sqrt{n^{\prime }+1}C_{lmn}^{\ast \left( 1\right) l^{\prime
}m^{\prime }n^{\prime }+1}C_{lmn}^{\left( 1\right) l^{\prime }m^{\prime
}n^{\prime }}+\sqrt{n^{\prime }}C_{lmn}^{\ast \left( 1\right) l^{\prime
}m^{\prime }n^{\prime }-1}C_{lmn}^{\left( 1\right) l^{\prime }m^{\prime
}n^{\prime }}\right) \\
&=&0.
\end{eqnarray*}%
Thus, for $f\left( \eta\right) =\eta,$ the uncertainty relations (\ref{20}) are:%
\begin{equation}
\left( \bigtriangleup q^{1}\right) _{\Psi }\left( \bigtriangleup
q^{2}\right) _{\Psi }\geq O\left( \theta ^{4}\right) ,  \label{44}
\end{equation}%
i.e., up to high enough order in $\theta $ nonlocality does not appear. So,
we can see that noncommutativity does not always lead to nonlocality.
However, the energy spectrum in this case changes, the first nontrivial
correction being:%
\begin{equation}
E_{lm0}^{\left( 2\right) }=\frac{\theta ^{2}}{8\omega _{\eta}}\left[ \omega
_{1}^{3}\left( l+\frac{1}{2}\right) +\omega _{2}^{3}\left( m+\frac{1}{2}%
\right) \right] .  \label{45}
\end{equation}
Despite the absence of nonlocality, the dynamical noncommutativity with $%
f\left( \eta\right) =\eta$ can contribute to other physical quantities.

For $f\left( \eta\right) =\eta^{2}$ in the first order in $\theta $ one has:%
\begin{equation}
\left( \bigtriangleup q^{1}\right) _{\Psi }\left( \bigtriangleup
q^{2}\right) _{\Psi }\geq \frac{\theta }{2\omega _{\eta}}\left( n+\frac{1}{2}%
\right) +O\left( \theta ^{3}\right) .  \label{46}
\end{equation}%
Thus, for $n=0$ the right hand side of this expression is just a constant,
like in the case of canonical nonlocality but interesting effects can be
seen in the higher orders in $\theta $. The second order contribution to (%
\ref{43}) is:%
\begin{eqnarray}
&&\frac{1}{2\omega _{\eta}}\left[ \sqrt{\left( n+1\right) \left( n+2\right) }%
\left( C_{lmn}^{\ast \left( 2\right) lmn+2}+C_{lmn}^{\left( 2\right)
lmn+2}\right) \right.  \notag \\
&&+\left( 2n+1\right) \left( C_{lmn}^{\ast \left( 2\right)
lmn}+C_{lmn}^{\left( 2\right) lmn}\right) +\sqrt{n\left( n+1\right) }\left(
C_{lmn}^{\ast \left( 2\right) lmn-2}+C_{lmn}^{\left( 2\right) lmn-2}\right)
\notag \\
&&+\underset{l^{\prime }m^{\prime }n^{\prime }}{\sum }\left( \sqrt{\left(
n^{\prime }+1\right) \left( n^{\prime }+2\right) }C_{lmn}^{\ast \left(
1\right) l^{\prime }m^{\prime }n^{\prime }+2}C_{lmn}^{\left( 1\right)
l^{\prime }m^{\prime }n^{\prime }}\right.  \notag \\
&&+\left. \left. \left( 2n^{\prime }+1\right) C_{lmn}^{\ast \left( 1\right)
l^{\prime }m^{\prime }n^{\prime }}C_{lmn}^{\left( 1\right) l^{\prime
}m^{\prime }n^{\prime }}+\sqrt{n^{\prime }\left( n^{\prime }+1\right) }%
C_{lmn}^{\ast \left( 1\right) l^{\prime }m^{\prime }n^{\prime
}-2}C_{lmn}^{\left( 1\right) l^{\prime }m^{\prime }n^{\prime }}\right) %
\right] .  \notag
\end{eqnarray}%
Now, we put $n=0$ and retain only the second order of the ratio $\omega
_{i}/\omega _{\eta}$, we came to
\begin{equation}
\frac{\theta ^{2}}{16\omega _{1}\omega _{2}\omega _{r}^{2}}\frac{\left(
\omega _{1}^{3}-\omega _{2}^{3}\right) ^{2}}{\left( \omega _{1}+\omega
_{2}\right) ^{2}}+\frac{\theta ^{2}\left( 2lm+l+m\right) }{16\omega
_{1}\omega _{2}\omega _{r}^{2}}\left[ \frac{\left( \omega _{1}^{3}-\omega
_{2}^{3}\right) ^{2}}{\left( \omega _{1}+\omega _{2}\right) ^{2}}+\frac{%
\left( \omega _{1}^{3}+\omega _{2}^{3}\right) ^{2}}{\left( \omega
_{1}-\omega _{2}\right) ^{2}}\right] .  \label{48}
\end{equation}

Thus, for the choice $f\left( \eta\right) =\eta^{2}$ the uncertainty relations (%
\ref{20}) are:%
\begin{eqnarray}
\left( \bigtriangleup q^{1}\right) _{\Psi }\left( \bigtriangleup
q^{2}\right) _{\Psi } &\geq &\frac{\theta }{4\omega _{\eta}}+\frac{\theta ^{3}}{%
32\omega _{1}\omega _{2}\omega _{\eta}^{2}}\frac{\left( \omega _{1}^{3}-\omega
_{2}^{3}\right) ^{2}}{\left( \omega _{1}+\omega _{2}\right) ^{2}}  \label{49}
\\
&&+\frac{\theta ^{3}\left( 2lm+l+m\right) }{32\omega _{1}\omega _{2}\omega
_{\eta}^{2}}\left[ \frac{\left( \omega _{1}^{3}-\omega _{2}^{3}\right) ^{2}}{%
\left( \omega _{1}+\omega _{2}\right) ^{2}}+\frac{\left( \omega
_{1}^{3}+\omega _{2}^{3}\right) ^{2}}{\left( \omega _{1}-\omega _{2}\right)
^{2}}\right] + O(\theta^{4}).  \notag
\end{eqnarray}%
In the third order in $\theta $ nonlocality depends on the quantum numbers $%
l $ and $m$, which determine the zero order energy (\ref{24}). We have
obtained the energy dependent nonlocality, the more the energy of the
system, the more the nonlocality (\ref{49}). Moreover, this contribution is
due to the physical degrees of freedom, which correspond to the coordinates $%
q^{1}$ and $q^{2}$. However, it should be noted that the effect is quite
small, comparable to the canonical noncommutativity, since the contribution
to the nonlocality is of the $\theta ^{3}$ order.

It should be noted that the energy dependent nonlocality also appears in other models of
noncommutativity, like kappa-Minkowski space and doubly special relativity; see for review \cite{JKG}.
However, it happens in a different manner, through modifications of energy-momentum dispersion relations
or energy-dependent generalizations of the spacetime metric \cite{AK}.

The first nontrivial correction to the energy is:%
\begin{eqnarray}
E_{lm0}^{\left( 2\right) } &=&\frac{\theta ^{2}}{32\omega _{1}\omega
_{2}\omega _{\eta}^{2}}\left[ \frac{\left( \omega _{1}^{3}+\omega
_{2}^{3}\right) ^{2}\left( l-m\right) }{\omega _{1}-\omega _{2}}-\frac{%
\left( \omega _{1}^{3}-\omega _{2}^{3}\right) ^{2}\left( l+m+1\right) }{%
\omega _{1}+\omega _{2}}\right]  \label{50} \\
&&+\frac{3\theta ^{2}}{64\omega _{\eta}^{2}}\left[ \omega _{1}^{3}\left( l+%
\frac{1}{2}\right) +\omega _{2}^{3}\left( m+\frac{1}{2}\right) \right] .
\notag
\end{eqnarray}%
Note that this quantity is of the order $\left( \omega _{i}/\omega
_{\eta}\right) ^{2}$, while the correction to the energy (\ref{45}) is of the
order $\omega _{i}/\omega _{\eta}$.

Note that if we choose $f\left( \eta\right) =\eta^{2}-1/\omega _{\eta}$, the
expression (\ref{49}) will have no contribution in the first order in $%
\theta $, first nontrivial contribution will appear only in $\theta
^{3}$ order. That is in this case the contribution of the physical degrees of
freedom to nonlocality is significant. Thus, by choosing $f\left( \eta\right) $
adequately we may control nonlocality, accordingly to physical
considerations.

\section{Conclusion}

In conclusion we note that the proposed model of noncommutativity is in fact
a generalization of canonical noncommutativity $\left[ \hat{x}^{i},\hat{x}%
^{j}\right] =i\theta ^{ij},$ which can be obtained just by taking $f\left(
\eta\right) =const$. However, dynamical noncommutativity provides much more
freedom in the description of the physical effects caused by
noncommutativity, connected with the presence of one more degree of freedom.
One interesting possibility would be to consider this degree of freedom as an
effective gravitational degree of freedom, which may correspond to a
compact extra dimension, like in Kaluza - Klein theory of gravity, or to an
extended extra dimension, like in Randall - Sundrum model.

We have discussed two different choices of theinteraction, i.e., the
function $f\left( \eta\right) $. In the first case, $f\left( \eta\right) =\eta$, the
nonlocality does not show up to quite higher orders in $\theta $, see (\ref%
{44}), while in the second one, $f\left( \eta\right) =\eta^{2}$, nonlocality
appears already in the first order in $\theta $, see (\ref{46}) and (\ref{49}%
). Thus, choosing the functions $V\left( r\right) $ and $f\left( r\right) $
from some physical considerations we are able to control noncommutativity,
nonlocality and its contribution to physical quantities like the energy
spectrum, etc.

\section*{Acknowledgements}

We are gratefull to Prof. H.O. Girotti and to Dr. F.S. Bemfica for very
usefull discussions. V.G.K. thanks the Institute of Physics of Universidade
Federal do Rio Grande do Sul for hospitality. V.G.K. acknowledges FAPESP
for support. M.G. and A.J.S. thank FAPESP and CNPq for partial support.

\end{document}